\documentclass[twoside,fleqn]{article}
\usepackage{espcrc2}
\usepackage{amssymb,amsmath}

\usepackage{graphicx}
\DeclareGraphicsExtensions{.jpg,.jpeg,.eps.ps}

\newcommand{\bra}{\langle}
\newcommand{\ket}{\rangle}

\newcommand{\order}{\mathcal{O}}

\newcommand{\psibar}{\bar\psi}
\newlength{\colw}
\newcommand{\calltoall}{\cite{Foley:2005ac}}
\newcommand{\caria}{\cite{Foley:2004jf}}

\title{Mesons at high temperature in $N_f=2$ QCD\thanks{Talk by
    JIS at Workshop on Computational Hadron Physics, University of
    Cyprus, 14--17 September 2005.}}

\author{Gert Aarts\address[Swan]{Department of Physics, University of
    Wales Swansea, Singleton Park, Swansea SA2 8PP, Wales, UK},
  Chris Allton\addressmark[Swan],
  Richie Morrin\address[TCD]{School of Mathematics, 
    Trinity College, Dublin 2, Ireland}, Alan \'O
    Cais\addressmark[TCD], Mehmet Bu\u{g}rahan Oktay\addressmark[TCD],
    Mike Peardon\addressmark[TCD], Jon-Ivar Skullerud\addressmark[TCD]
}

\begin{document}
\makeatletter \@mathmargin = 0pt \makeatother
\bibliographystyle{h-elsevier3}

\begin{abstract}
We report first results for spectral functions of charmonium in
2-flavour QCD.  The spectral functions are determined from vector and
pseudoscalar correlators on a dynamical, anisotropic lattice.
$J/\psi$ and $\eta_c$ are found to survive well into the deconfined
phase before melting away at $T\lesssim2T_c$.  Current systematic
uncertainties prevent us from drawing any definite conclusions at this
stage.
\end{abstract}

\maketitle

\section{INTRODUCTION}

The properties of hadrons or hadronic resonances above the
deconfinement transition is a subject at the heart of the current
experimental programme at RHIC, where hadronic signals are used to
obtain information about the state of matter inside the fireball.  The
questions of interest include the issue of which hadrons survive as
bound states in the quark--gluon plasma, and up to which temperature;
as well as the transport properties of light and heavy quarks in the
plasma.

These properties are all encoded in the spectral functions
$\rho(\omega,\vec{p})$, which are related to the imaginary-time
correlator $G_\Gamma(\tau,\vec{p})$ according to
\begin{equation}
G_\Gamma(\tau,\vec{p}) = \frac{1}{2\pi}
\int_0^\infty\rho_\Gamma(\omega,\vec{p})K(\tau,\omega)d\omega\,,
\label{eq:spectral}
\end{equation}
where the subscript $\Gamma$ correspond to the different quantum
numbers.  The kernel $K$ is given by
\begin{equation}
\begin{split}
K(\tau,\omega) &= \frac{\cosh[\omega(\tau-1/2T)]}{\sinh(\omega/2T)} \\
 &= e^{\omega\tau}n_B(\omega) + e^{-\omega\tau}[1+n_B(\omega)]\,.
\end{split}
\label{eq:kernel}
\end{equation}
and $n_B$ is the Bose--Einstein distribution function.

The spectral function can be extracted from the lattice correlators
$G(\tau)$ using the Maximum Entropy Method (MEM)
\cite{Asakawa:2000tr}.  For this to work and give reliable results, it
is necessary to have a sufficient number of points in the euclidean
time direction.  This will be prohibitively expensive, especially when
dynamical quarks are included, unless anisotropic lattices are used,
with a temporal lattice spacing much smaller than the spatial lattice
spacing.

We will here focus on the charmonium S-wave states $\eta_c$ and
$J/\psi$ at zero momentum, which have attracted much attention
following the suggestion \cite{Matsui:1986dk} that $J/\psi$
suppression could be a probe of deconfinement.  Potential model
calculations using the heavy quark free energy have tended to support
this picture.  However, previous simulations in the quenched
approximation \cite{Umeda:2002vr,Asakawa:2003re,Datta:2003ww} indicate
that contrary to this, $J/\psi$ may survive up to temperatures as high
as $1.5-2T_c$.  Recently, potential model calculations using the
internal energy of the heavy-quark pair have reached the same
conclusion, and using the most recent lattice data
\cite{Kaczmarek:2005gi} these models indicate a qualitatively similar
picture in the case of $N_f=2$ QCD
\cite{Kaczmarek:2005gi,Wong:2005be}.

In this study we attempt to determine directly the spectral functions
of charmonium in 2-flavour QCD using anisotropic lattices and the
Maximum Entropy Method.

\section{SIMULATION DETAILS}

We use the Two-plaquette Symanzik Improved gauge action
\cite{Morningstar:1999dh} and the fine-Wilson, coarse-Hamber-Wu
fermion action \cite{Foley:2004jf} with stout-link smearing
\cite{Morningstar:2003gk}.  The process of tuning the action
parameters, and the parameters used, are described in more detail in
\cite{Morrin:2005tc,Ryan:2005yy}.  The parameters correspond to a
spatial lattice spacing $a_s\approx0.2$fm with an anisotropy
$\xi=a_s/a_t\approx6$. The sea quark mass corresponds to
$m_\pi/m_\rho\approx0.55$.

These actions are designed for large anisotropies, and a quenched
study \caria\ found that no mass-dependent tuning of the
quark anisotropy was necessary up to valence quark masses well beyond
charm.  This appears no longer to be the case for dynamical quarks;
indeed, for the parameters used here the anisotropy determined from
the pion dispersion relation was found to be $\xi_q^{\ell}\sim6.4$, in
rough agreement with the gluon anisotropy, while the anisotropy from
the charmonium dispersion relation was found to be $\xi_q^h\sim8$.
This issue is still under investigation.

For this preliminary study we have used a spatial volume of
$N_s^3=8^3$ and generated 100 configurations, sampled every 10 HMC
trajectories, for $N_t=48,32,24$ and 16.  $N_t=32$ corresponds to a
temperature close to the pseudocritical temperature $T_c$, although
larger lattices will be needed to determine this with any precision.
We have computed charmonium correlators in the pseudoscalar ($\eta_c$)
and vector ($J/\psi$) channels with bare charm quark mass
$a_tm_c=0.1$.  The charmonium spectrum in the hadronic phase is
presented in \cite{Ryan:2005yy,Juge:2005nr}.  In this study we have
used local (unsmeared) operators,
\begin{equation}
G_\Gamma(\tau) =
\sum_{\vec{x},\vec{y},t}\bra
M^\dagger_\Gamma(\vec{x},t)M_\Gamma(\vec{y},t+\tau)\ket\,,
\end{equation}
where
\begin{equation}
M_\Gamma(\vec{x},t)=\psibar(\vec{x},t)\Gamma\psi(\vec{x},t)\,,\quad
\Gamma=\gamma_5,\gamma_i\,.
\end{equation}
All-to-all propagators \calltoall\
have been used to improve the signal and sample information from the
entire lattice.  The propagators were constructed with no eigenvectors
and two noise vectors diluted in time, colour and even/odd in space.

The MEM analysis has been performed with the continuum free spectral
function $\omega^2$ as default model, using the euclidean correlators
in a time window starting at  $\tau=2$, and cutting off the energy
integral in (\ref{eq:spectral}) at $a_t\omega_{\text{max}}=6$.

\section{RESULTS}

\begin{figure}
\includegraphics*[width=\colw]{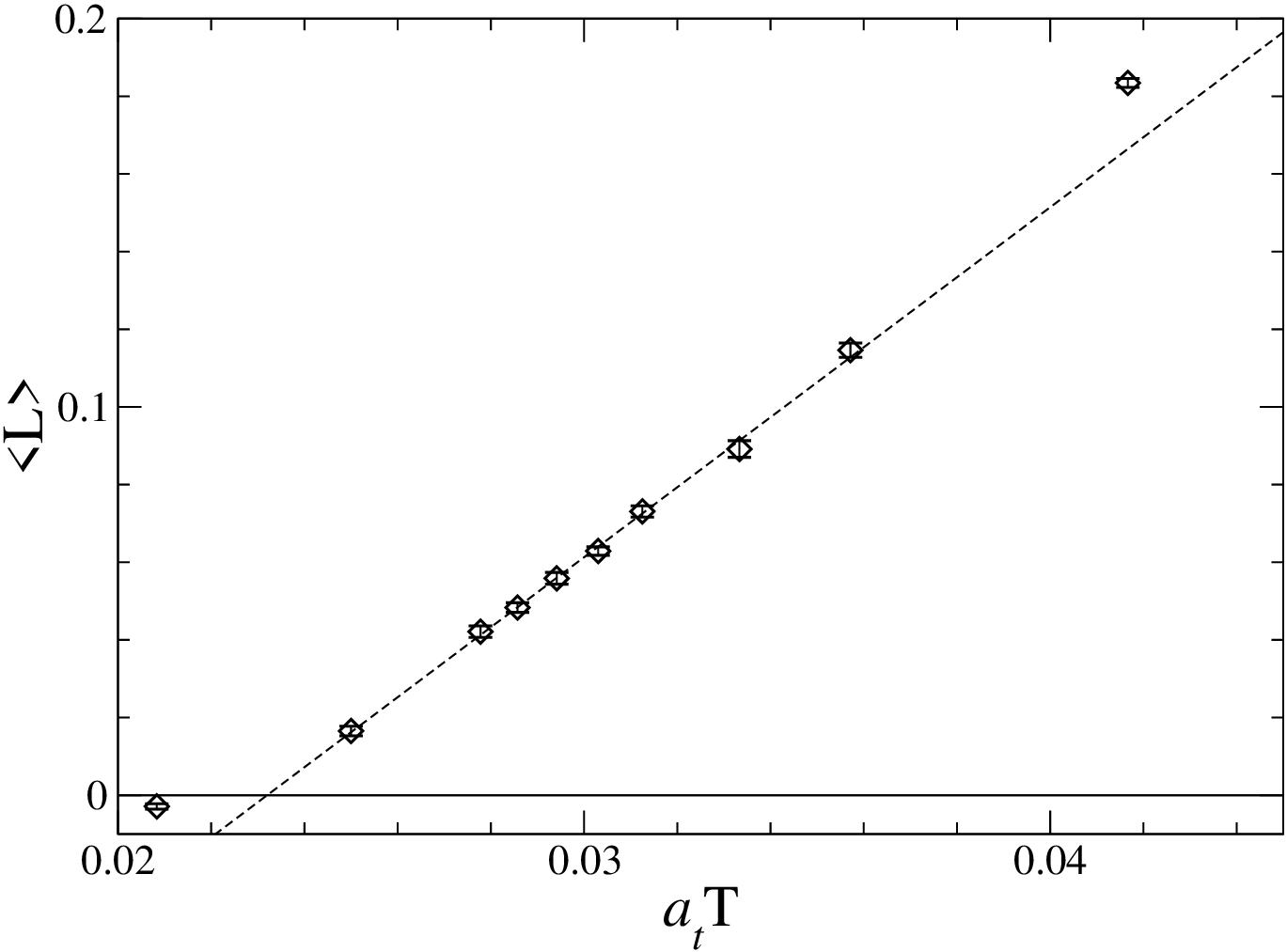}
\caption{Average Polyakov loop as a function of temperature.}
\label{fig:polyakov}
\end{figure}

In an attempt to locate the pseudocritical temperature on these
lattices, we performed simulations varying $N_t$ in the range 28--40,
measuring the real part of the Polyakov loop $\bra L\ket$ and its
susceptibility.  The results for the Polyakov loop are shown in
fig.~\ref{fig:polyakov}.
As can be seen from the figure, the Polyakov line in the transition
region follows a linear behaviour in temperature, and therefore no
pseudocritical temperature can be determined on these lattices.  It
will be necessary to use larger lattices to determine $T_c$.  For this
reason, and because of the uncertainties regarding the anisotropy, we
will refrain from quoting results in terms of $T_c$.

\begin{figure}[t]
\includegraphics*[width=0.45\textwidth]{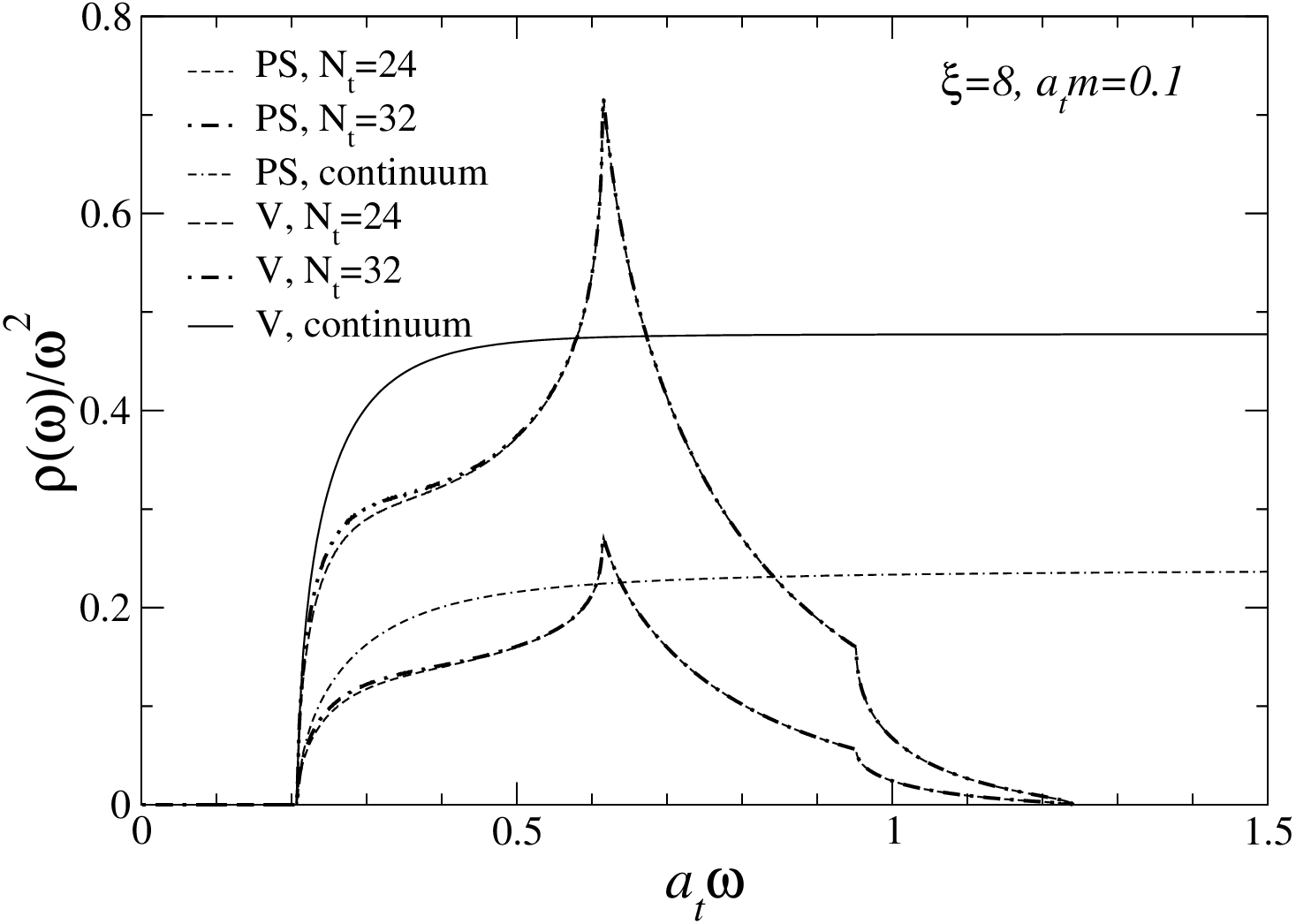}
\vspace{-0.7cm}
\caption{Free-fermion spectral functions in the pseudoscalar
  and vector channels, for bare quark mass $a_tm_0=0.1$.}
\label{fig:free}
\end{figure}
The free-fermion spectral functions are shown in fig.~\ref{fig:free},
for $N_t=32$ and 24, and in the continuum.  The most striking feature
of the lattice free spectral functions is the spike at
$a_t\omega\approx0.6$. This means that lattice artefacts are big at
this point, and results for spectral functions cannot be trusted in
this region.  The lattice functions undershooting the continuum curve
appears to be a mass-dependent, $\order(a_tm)$ or $\order(a_s^2m)$,
effect, evidenced by the effect being much larger at
$a_tm=0.2$~\cite{Morrin:2005zq}.

\begin{figure}
\includegraphics*[width=\colw]{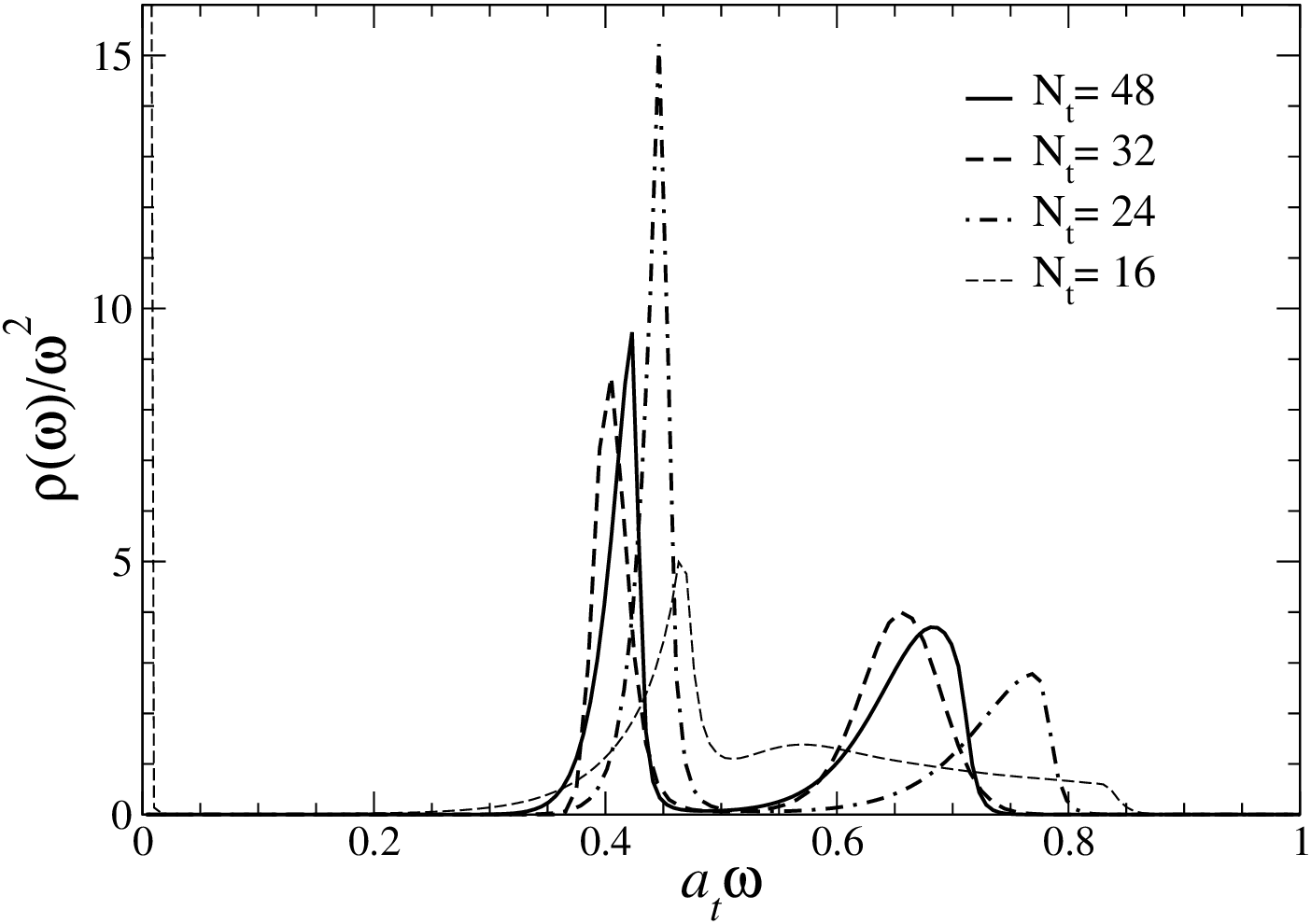}
\vspace{-0.7cm}
\caption{Pseudoscalar ($\eta_c$) spectral function for different
  temperatures.}
\label{fig:spf-p5}
\end{figure}
\begin{figure}
\includegraphics*[width=\colw]{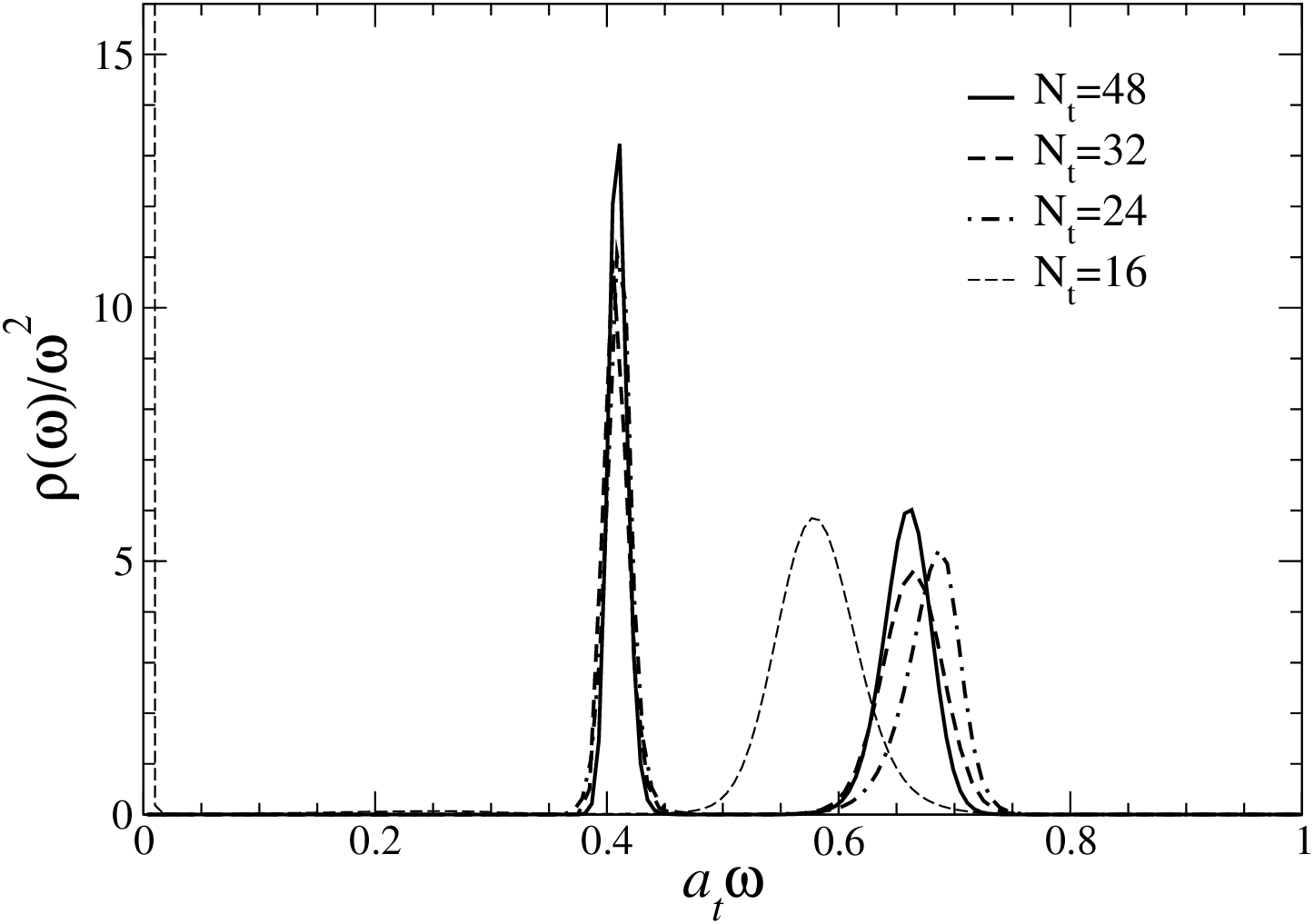}
\vspace{-0.7cm}
\caption{Vector ($J/\psi$) spectral function for different temperatures.}
\label{fig:spf-vi}
\end{figure}

The spectral functions obtained from the MEM analysis are shown in
figs.~\ref{fig:spf-p5} and \ref{fig:spf-vi} for $\eta_c$ and $J/\psi$
respectively.  The main peak position for $N_t=48$ agrees within
errors with the mass obtained for the respective particles on the same
lattices, using a variational basis of smeared operators
\cite{Ryan:2005yy,Juge:2005nr}.
The second peak coincides with the cusp in the free spectral function
in fig.~\ref{fig:free}, and it can therefore be concluded that this is
primarily a lattice artefact.  The radially excited states cannot be
resolved with our present statistics.

The results in figs.~\ref{fig:spf-p5} and \ref{fig:spf-vi} indicate
that the 1S states survive in the medium up to well beyond the
deconfinement temperature, finally melting away at $T\lesssim2T_c$.
This is in qualitative agreement with recent potential model results
using the static quark--antiquark internal energy
\cite{Kaczmarek:2005gi}.  Using the colour singlet internal energy
$U_1$ yields a dissociation temperature of $T_{\text{dis}}\sim2T_c$,
while a potential constructed to exclude the gluon internal energy 
\cite{Wong:2004zr} yields $T_{\text{dis}}\sim1.4T_c$
\cite{Wong:2005be}.  Given the uncertainties in this calculation, both
these results are consistent with the present data.

It is not clear whether the apparent stronger binding in $\eta_c$ for
$N_t=24$ is significant, in particular given that no such effect is
observed for $J/\psi$.  Higher statistics will be needed to resolve
this.  If it is confirmed, it might be in line with potential model
calculations using $U_1$, which is strongly peaked around $T_c$.

\section{OUTLOOK}

Using an anisotropic lattice, we have performed the first calculation
of charmonium spectral functions in 2-flavour QCD.  The results appear
to confirm the picture emerging from quenched simulations, that
$J/\psi$ and $\eta_c$ survive until well into the deconfined phase.

These simulations have been performed with parameters that are not
fully tuned, giving rise to significant systematic uncertainties.  A
particular issue is the tuning of the charm quark anisotropy.  We are
in the process of obtaining a fully tuned parameter set, and
simulations at these parameters will be carried out in the very near
future.  Firm conclusions will have to await these simulations.

The small lattice volume --- only $(1.6\text{fm})^3$ --- is also a
major source of systematic uncertainty, in particular as it has
prevented a determination of the pseudocritical temperature.  Future
simulations will be carried out on larger spatial volumes.  It will
also be important to increase the statistics, in order to resolve
excited states and disentangle the effects of thermal width and finite
statistics.  We will also be carrying out a systematic study of the
effects of using different default models, including the free lattice
spectral functions, and different time and energy ranges in the MEM
analysis.  Initial indications are that our results are relatively
robust against such changes.

The coarse spatial lattice is an issue in that it gives rise to
significant, mass-dependent, lattice artefacts as shown in
fig.~\ref{fig:free}.  It will ultimately be necessary to repeat the
calculation on a finer lattice; this will however require a new
nonperturbative tuning and is therefore not on the immediate horizon.
Some information about lattice spacing effects may be gleaned from
quenched studies using the same action, which are considerably cheaper
as the quark and gluon anisotropies can be tuned independently.

We are planning to compute the spectral functions also at non-zero
momentum \cite{Aarts:2005hg}, which contain additional information not
found at zero momentum, and which may relate more directly to
experimental data taken at non-zero momentum.

We will also study light vector meson correlators at zero and non-zero
momentum, which can be related to dilepton production rates in the
plasma.  Finite lattice spacing effects are expected to be less of a
problem in this case, so the current lattices are likely to be
suitable for this purpose.  This will proceed once the fully tuned
parameter set is ready.

\section*{Acknowledgments}

We thank the organisers of {\em Workshop on Computational Hadron
Physics} for a pleasant and interesting workshop.  This work has been
supported by the IRCSET Embark Initiative award SC/03/393Y, SFI grant
04/BRG/P0275 and the IITAC PRTLI initiative.  We wish to thank Jimmy
Juge, Sin{\'e}ad Ryan and Simon Hands for stimulating and fruitful discussions.

\bibliography{trinlat,hot}
\end{document}